\documentclass{elsart}

\usepackage{natbib}
\usepackage{epsfig}
\usepackage{amssymb}

\begin{document}

\begin{frontmatter}

\title{Partially ionized atmospheres of neutron stars 
with strong magnetic fields}

\author[ioffe]{A.Y. Potekhin\corauthref{cor}},
\corauth[cor]{Corresponding author. Tel.: +7-812-2479180;
fax: +7-812-5504890.}
\ead{palex@astro.ioffe.ru}
\author[cornell]{Dong Lai},
\author[ens]{G. Chabrier},
\author[kavli]{W.C.G. Ho\thanksref{fellow}}
\thanks[fellow]{Hubble fellow.}

\address[ioffe]{Ioffe Physico-Technical Institute,
194021 St.~Petersburg, Russia}
\address[cornell]{
 Department of Astronomy,
 Cornell University,
 Ithaca, NY 14853, USA}
\address[ens]{Ecole Normale Sup\'erieure de Lyon,
  CRAL (UMR CNRS No.\ 5574), \\
 46 all\'ee d'Italie, 
69364 Lyon, 
 France}
\address[kavli]{Kavli Institute for Particle Astrophysics and Cosmology,
Stanford University,
Stanford, CA 94309, USA}

\begin{abstract}
We construct hydrogen atmosphere models for strongly
magnetized neutron stars in thermodynamic equilibrium,
taking into account partial ionization. The presence of bound states
affects
the equation of state, absorption coefficients, and
polarizability tensor of a strongly magnetized plasma.
Therefore the partial ionization influences the
polarization vectors and opacities of normal electromagnetic waves,
and thus the spectra of outgoing radiation.
Here we review a model suitable for
the most typical neutron-star atmospheres and
focus on the problems
that remain to be solved for its extension
to other atmospheric parameters.
\end{abstract}

\begin{keyword}
stellar atmospheres \sep neutron stars \sep strong magnetic fields
\end{keyword}

\end{frontmatter}

\section{Introduction}

Thermal emission from neutron stars (NSs) may potentially 
be used to directly measure the
NS surface magnetic field, temperature, and composition,
achieve a more complete understanding of the evolution 
of the NSs, and constrain the properties of matter and physical
processes
under extreme conditions. It was realized long ago
that a NS
atmosphere model should properly include a strong magnetic field
and partial ionization \citep[see, e.g.,][for an early review]{Pavlov-ea95}.
Models of \emph{fully ionized} NS atmospheres with strong magnetic fields
were constructed by several research groups 
\citep[e.g.,][and references therein]{Shib92,Zane00,HoLai02}.
The most recent papers highlighted the effects that
may be important for the atmospheres of magnetars: the ion cyclotron feature
\citep{HoLai,Zane01} and vacuum polarization effects, 
including a conversion of the normal modes of radiation propagating in the 
magnetized atmosphere \citep{HoLai02,LaiHo02,LaiHo03}.

Early considerations of \emph{partial ionization}
in the magnetized NS atmospheres 
(\citealt{Miller92,RRM}; also reviewed briefly by \citealt{ZP02}) 
were
not very reliable because of oversimplified treatments of
atomic physics and nonideal plasma effects in strong magnetic fields.
At the typical NS atmosphere parameters, 
the effects of thermal motion of the bound species are important.
In the 1990s, binding energies
and radiative transition rates with allowance
for the motion effects in strong magnetic fields 
have been calculated for the H atom \citep{P94,PP97}.
Recently these atomic data
 have been implemented in calculations
of thermodynamic functions \citep{PCS99,PC03,PC04}, 
radiative opacities \citep{PC03,PC04},
and spectra \citep{hoetal}
of the partially ionized H atmospheres of the NSs.
Some results have been presented at the previous {\sc Cospar} meeting
\citep{Ho-COSPAR}. Now our atmosphere model has been complemented by
the effects of the bound states on the polarization properties of 
the strongly magnetized plasma \citep{KK}.
Below we briefly summarize the results that allow us to calculate
realistic X-ray spectra of thermal radiation from hydrogen NS
atmospheres with magnetic fields
$B\sim10^{12}-10^{14}$~G and 
effective temperatures $T\gtrsim10^{5.5}$~K,
and outline the problems that remain to be
addressed at other atmospheric parameters and compositions.

\section{The atmosphere model}

We use the equation of state (EOS) for H in strong magnetic fields
\citep{PCS99} based on
the free-energy minimization method, which ensures the thermodynamic
consistency and allows one to determine 
number fractions of chemical species, required for opacity calculations. 
The model takes into account all
available theoretical results on the
moving H atoms and nonideal Coulomb plasmas in the magnetic fields. 
This EOS has been tabulated and employed for calculation of
opacities for astrophysical
use \citep{PC03,PC04}.

\begin{figure}[t]
\begin{center}
\epsfxsize=\textwidth
\epsffile{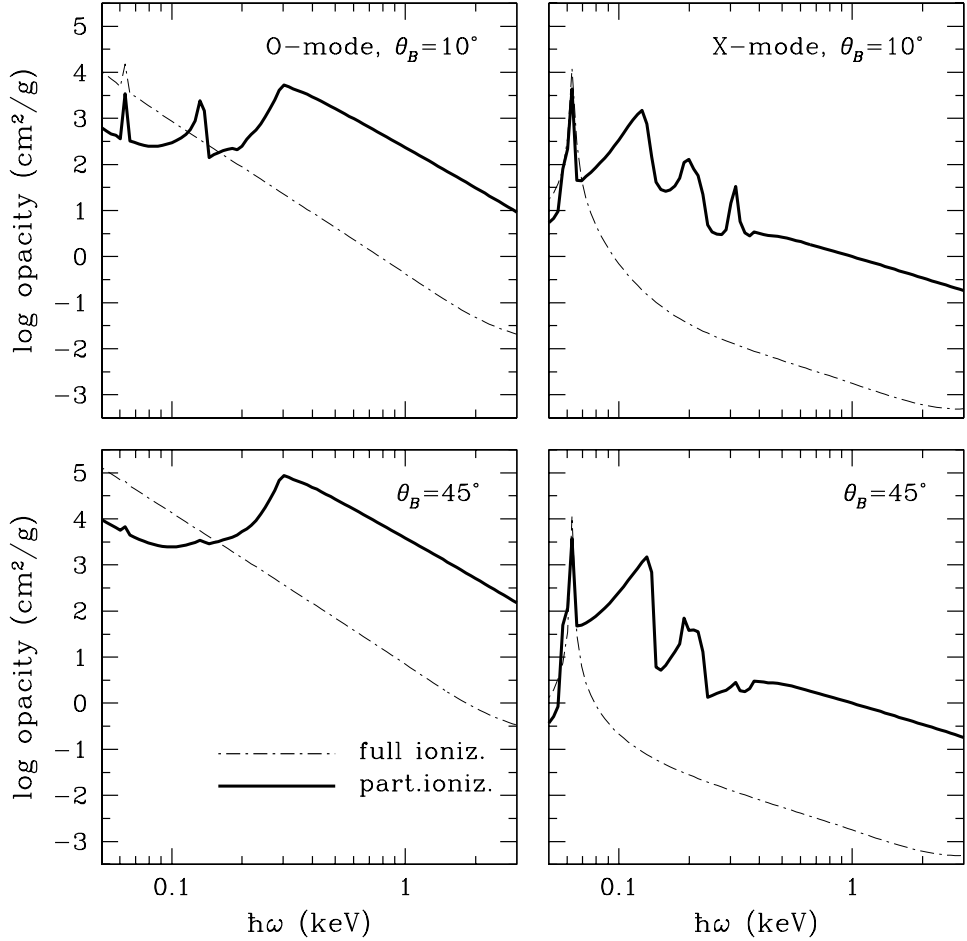}
\caption{Opacities $\kappa_j$ (Eq.~[\ref{opac}]) in the O-mode (left panels)
and X-mode (right panels)
versus
photon energy in the hydrogen plasma at $B=10^{13}$~G,
$T=10^{5.5}$~K, and $\rho=1$ g cm$^{-3}$, 
for $\theta_B=10^\circ$ (upper panels)
and $45^\circ$ (lower panels).
Solid lines: a self-consistent calculation
for a partially ionized plasma (70\% of neutrals);
dot-dashed lines: the model of full ionization.
\label{fig-ang13_0}
}
\end{center}
\end{figure}

It is well known that under typical
conditions (e.g., far from the resonances) radiation propagates in a
magnetized plasma in the form of two normal modes,
called the extraordinary (X) and the ordinary (O). 
The opacity in the mode $j$ ($j=$X,O)
depends  on the
photon frequency $\omega$, magnetic field $B$, density $\rho$,
temperature $T$, and the angle
$\theta_B$ between the magnetic field and propagation direction.
It can be written as
\begin{equation}
  \kappa_j(\omega,\theta_B) = \sum_{\alpha=-1}^1 \!\!
   |e_\alpha^j(\omega,\theta_B)|^2 \,\hat\kappa_\alpha(\omega),
\label{opac}
\end{equation}
where $e_\alpha^j$ ($\alpha=-1,0,1$) are the cyclic coordinates
of the polarization vectors 
of the normal modes, and the quantities
$\hat\kappa_\alpha$ ($\alpha=-1,0,1$)
do not depend on $\theta_B$.
\citet{PC03} calculated $\hat\kappa_\alpha(\omega)$
and evaluated $\kappa_j(\omega,\theta_B)$
 using the polarization vectors of the normal modes in
the fully ionized plasma,
Such a calculation (dubbed ``hybrid'' in \citealt{KK})
was employed in our previous model of partially ionized 
hydrogen
atmospheres of the NSs with strong magnetic fields 
\citep{hoetal,Ho-COSPAR}. 

In the new model
\citep{KK}, we take into account the 
influence of the bound species on the polarization vectors of 
the normal modes, making use of the Kramers-Kronig relation\footnote{%
Previously this relation has been used by \citet{BulikPavlov}
 for a neutral gas of H atoms in a strong magnetic field.}
between the imaginary and real parts of the plasma polarizability. Thus
the calculation of the polarization vectors and opacities
of the normal modes has become self-consistent. The 
calculations of thermal spectra of the NSs 
show that such self-consistent
treatment is necessary if the number fraction of
the bound states exceeds several percent.
In Fig.~\ref{fig-ang13_0} we compare 
radiative opacities calculated with and without allowance for 
the bound species 
for one particular set of plasma parameters, typical near
the radiative surface of a moderately cool neutron star
with magnetic field $B=10^{13}$~G, for two
$\theta_B$ values. In the case shown in the figure,
the neutral fraction is 70\%, thus the self-consistent treatment
of the opacities is important.

\section{Conclusion and unsolved problems}

The constructed atmosphere models allow us to calculate realistic spectra of
thermal X-ray radiation from H atmospheres of
the NSs with $10^{12}\mbox{ G}\lesssim B \lesssim 10^{14}\mbox{ G}$ 
and $T\gtrsim10^{5.5}$~K. Examples of these spectra are presented elsewhere
\citep{KK}.
There remain the following 
unsolved problems that
prevent us from making
 similarly reliable calculations beyond these limits.

{1.} Although the H EOS and opacities have been calculated for $B$
up to $10^{15}$~G and implemented in the atmosphere
model \citep{hoetal,Ho-COSPAR}, the calculated spectra  at
$B\gtrsim10^{14}$~G depend on the adopted model of mode conversion due to
the vacuum resonance and on description of propagation of photons with
frequencies below the plasma frequency.  Both these problems have not been
definitely solved. Their solution is also  important for modeling the
low-frequency (UV and optical) tail of the spectrum.

{2.} At lower $T$ or higher $B$, H atoms recombine in
H$_n$ molecules and eventually form the condensed phase 
\citep[][and references therein]{Lai-RMP}.
Corresponding quantum-mechanical data are very incomplete.

{3.} At $10^9\mbox{ G}\lesssim B \lesssim 10^{11}$~G, 
transition rates of the moving H atom have
 not been calculated previously because of their
complexity. 
The first calculation of the energy spectrum
appropriate in this range of $B$
has been published when the present paper was in preparation
\citep{LozovikVolkov}.

{4.} At present it is not possible to calculate accurate
atmospheric spectra at $B\gtrsim10^{12}$~G for chemical elements
other than hydrogen, because of the importance of
the effects of finite nuclear mass in the strong field regime.
Apart from the H atom, these effects have been
calculated only for the He atom \emph{at rest} \citep{Hujaj03a,Hujaj03b}
and for the He$^+$ ion (at only one value of $B$, \citealt{BPV}).

{5.} A more rigorous treatment of radiative transfer in the atmosphere
requires solving the transfer equations of the Stokes parameters (see,
e.g., \citealt{LaiHo03} for the cases of fully ionized atmospheres).
However, since the nonorthogonal features of the modes due to neutral
species are  diminished by the center-of-mass motion, the effect is
expected to be small.

Finally, let us note that the atmosphere model presented here, together with
a model of radiation from the condensed magnetic surface
\citep{surfem}, has been successfully used for fitting the spectrum
of the isolated neutron star RX J1856.5$-$3754 \citep{Ho-RXJ}.

\textbf{Acknowledgements.}
A.P.\ acknowledges the hospitality
of the Astronomy Department of Cornell University
and the theoretical astrophysics group 
at the Ecole Normale Su\-p\'e\-r\-i\-eure de Lyon.
The work of A.P.\ is supported in part by RFBR grants 
02-02-17668 and 03-07-90200, 
and RLSS grant 1115.2003.2. 
D.L.\ is supported in part by NSF grant AST 0307252,
NASA grant NAG 5-12034,
and SAO grant TM4-5002X.
W.H.\ is supported by NASA through Hubble Fellowship grant
HF-01161.01-A awarded by STScI, which is operated by AURA, Inc.,
for NASA, under contract NAS 5-26555.

\end{document}